\documentclass[journal,twoside,web]{Classes/IEEEtran}
\usepackage{tmi}
\usepackage{cite}
\usepackage{amsmath,amssymb,amsfonts}
\usepackage{algorithmic}
\usepackage{booktabs}
\usepackage{graphicx}
\usepackage{textcomp}
\usepackage{stfloats}

\usepackage{multirow}
\usepackage{hyperref}

\begin{document}
\title{Bayesian Inference of Tissue Heterogeneity for Individualized Prediction of Glioma Growth}
\author{Baoshan Liang, 
Jingye Tan,
Luke Lozenski,
David A. Hormuth II,
Thomas E. Yankeelov,
Umberto Villa,
Danial Faghihi
%
\thanks{B. Liang is with 
the Mechanical and Aerospace Engineering Department, University at Buffalo, Buffalo, NY, USA
(e-mail: baoshanl@buffalo.edu).}
\thanks{J. Tan is with 
the Mechanical and Aerospace Engineering Department, University at Buffalo, Buffalo, NY, USA
(e-mail: jtan32@buffalo.edu).}
\thanks{L. Lozenski is with 
the Electrical and Systems Engineering Department, Washington University in St. Louis, St. Louis, MO, USA
(e-mail: ljlozenski@wustl.edu).}
\thanks{D. A. Hormuth II is with the
Oden Institute for Computational Engineering and Sciences, Livestrong Cancer Institutes, The University of Texas at Austin, Austin, TX, USA (e-mail: david.hormuth@austin.utexas.edu).}
\thanks{T. E. Yankeelov is with the
Oden Institute for Computational Engineering and Sciences, Livestrong Cancer Institutes, Departments of Biomedical Engineering, Diagnostic Medicine, and Oncology, The University of Texas at Austin, Austin, TX, USA, Department of Imaging Physics, The University of Texas MD Anderson Cancer Center, Houston, TX, USA (e-mail: thomas.yankeelov@utexas.edu).}
\thanks{U. Villa is with the
Oden Institute for Computational Engineering and Sciences, The University of Texas at Austin, Austin, TX, USA
(e-mail: uvilla@austin.utexas.edu).}
\thanks{D. Faghihi (corresponding author) is with 
the Mechanical and Aerospace Engineering Department, University at Buffalo, Buffalo, NY, USA
(e-mail: danialfa@buffalo.edu).}
}

\maketitle

\begin{abstract}

Reliably predicting the future spread of brain tumors using imaging data and on a subject-specific basis requires quantifying uncertainties in data, biophysical models of tumor growth, and spatial heterogeneity of tumor and host tissue. 
%
This work introduces a Bayesian framework to calibrate the spatial distribution of the parameters within a tumor growth model to quantitative magnetic resonance imaging (MRI) data and demonstrates its implementation in a pre-clinical model of glioma. The framework leverages an atlas-based brain segmentation of grey and white matter to establish subject-specific priors and tunable spatial dependencies of the model parameters in each region. 
%
Using this framework, the tumor-specific parameters are calibrated from quantitative MRI measurements early in the course of tumor development in four rats and used to predict the spatial development of the tumor at later times. The results suggest that the tumor model, calibrated by animal-specific imaging data at one time point, can accurately predict tumor shapes with a Dice coefficient $>$ 0.89. However, the reliability of the predicted volume and shape of tumors strongly relies on the number of earlier imaging time points used for calibrating the model.  
%
This study demonstrates, for the first time, the ability to determine the uncertainty in the inferred tissue heterogeneity and the model predicted tumor shape.

\end{abstract}

\begin{IEEEkeywords}
Computational oncology,
uncertainty quantification,
biophysical tumor model,
tumor shape prediction,
quantitative MRI.
\end{IEEEkeywords}


\section{Introduction}
\label{sec:introduction}
Glioblastoma is the most aggressive primary brain tumor and is characterized by highly invasive tumor cells that infiltrate the surrounding tissue \cite{dolecek2012}. Importantly, these diffusive tumor boundaries in glioblastoma cannot be correctly identified based on standard medical imaging techniques, contributing to suboptimal interventions and poor prognosis \cite{stupp2005}. However, reliable model predictions of a tumor’s spatial and temporal development can augment the information available in medical images, potentially enabling clinicians to design optimal therapeutic regimens for individual patients. 
%
%
In recent years, several biophysical models of tumor growth have been developed and calibrated using non-invasive imaging measurements to predict future tumor growth and treatment outcomes (see, e.g., \cite{lima2014hybrid, Hormuth2022, lipkova2019, faghihi2020, roque2017dce, rockne2015, hormuth2021towards}). Despite these advances, predictive modeling of tumor development requires addressing two critical challenges. 
The first is characterizing uncertainties in the model predictions to improve subject-specific treatment outcomes over the standard protocols. The sources of uncertainties in tumor model predictions are the inadequacy of models to capture the complexity of the underlying biological processes of tumor growth as well as the noise and scarcity of observational data on tumor development \cite{oden2016toward}. Quantitatively evaluating the uncertainty in predictions of tumor growth on a subject-specific basis has been gaining attention, and Bayesian calibration of biology-based tumor growth models with clinical and pre-clinical MRI data \cite{lima2017selection, lipkova2019, le2016mri} is a natural way to advance the field. 
The second challenge is characterizing the spatial heterogeneity of the tumor and host tissue that leads to irregular patterns of tumor spread with pronounced effects on the ability to deliver therapeutics \cite{sefidgar2014} and design optimal radiation plans \cite{adeberg2018}. 
Nevertheless, the estimation of uncertainty in computational models' prediction of tumor shape has not yet been addressed. The ability to reliably predict the development of individual tumor shapes would dramatically improve the clinical relevance of computational modeling of tumor growth and positively impact overall survival.


This contribution addresses these two challenges by developing a Bayesian framework for high-dimensional parameter inference. The ability of this framework to predict the future development of tumor shape is demonstrated using a standard reaction-diffusion tumor model \cite{harpold2007} and pre-clinical longitudinal MRI data of a murine model of glioma. The framework enables learning the spatial distributions of model parameters (corresponding to proliferation and random movement of tumor cells) in the brain domain from each animal’s imaging data while simultaneously quantifying the modeling error and measurement noise. These uncertainties are then propagated through the model to predict the tumor boundary at future time points with quantified uncertainties. 
An innovative feature of the proposed framework is accounting for the spatial correlation of the model parameter fields, which encode surrounding tissue information to drive the heterogeneous tumor expansion of the model. Moreover, the Bayesian inference is equipped with the subject-specific prior probability density of the parameters by developing an atlas-based segmentation technique to provide anatomical context (grey and white matter) of the brain.
Both features are shown to enhance the ability of the calibrated model to predict the spatial patterns of individual tumor development compared to other model calibration methods.  

\begin{figure*}[t]
    \centering
 \includegraphics[width=0.97\textwidth]{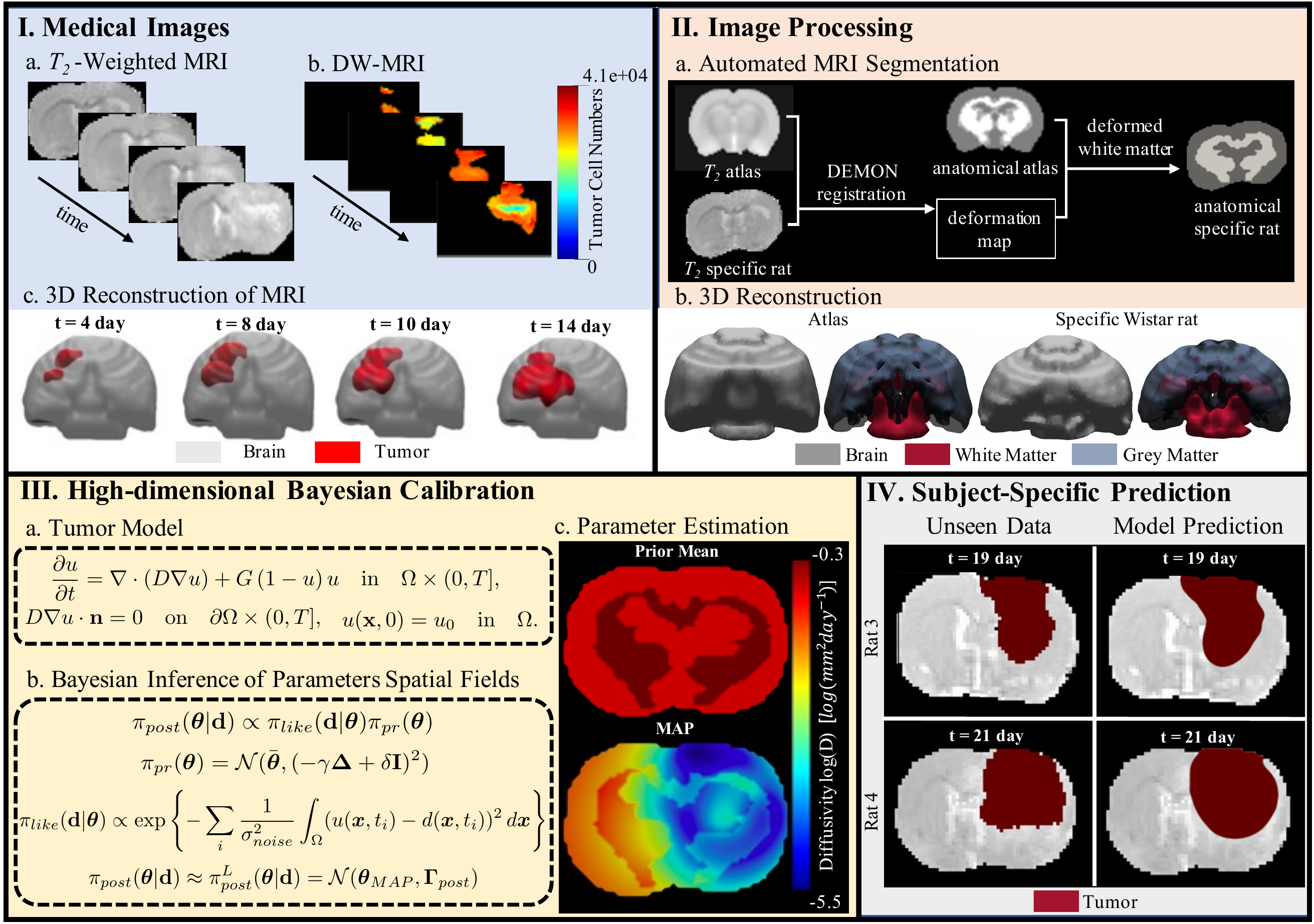}
 \vspace{-0.05in}
 \caption{
Overview of the high-dimensional Bayesian framework for subject-specific brain tumor prediction in Wistar rats. 
\textbf{I) MRI data} used for model calibration: (I.a) $T_2$-weighted MR images of the central slice of rats are used to define the simulation domain and segment the brain tissue. (I.b) DW-MRI measurements provide an estimate of the spatial distribution of tumor cells at different times that provide calibration data in terms of tumor volume fraction $d(\mathbf{x}, t_i)$, and (I.c) 3D reconstruction of the brain in coronal view using 2D slices of both imaging modalities showing tumor evolution in time. 
\textbf{II) Image processing} including (II.a) the atlas-based automated segmentation of the brain into grey and white matter, and (II.b) 3D reconstruction of the brain in the atlas and specific rat indicating grey and white matter. 
\textbf{III) High-dimensional Bayesian calibration} including (III.a) a mathematical model of tumor growth, which simulates the evolution of tumor volume fraction $u(\mathbf{x}, t_i)$ in animal anatomy, (III.b) Bayesian inference using subject-specific prior, and (III.c) estimation of model parameters as spatial fields from imaging data. 
\textbf{IV) Subject-specific prediction of tumor volume} at a future time point and comparison with the corresponding MRI data that was not used to calibrate the model.
 }
 \label{fig:overview}
 \vspace{-0.2in}
\end{figure*}

In the remainder of the paper, Section \ref{sec:methods} summarizes the data and image processing methods and introduces the Bayesian framework for image-driven inference of tumor model parameter fields. Section \ref{sec:results} presents the results, where the framework is applied to pre-clinical imaging data, and the model’s predictive ability is quantified. Finally, Section \ref{sec:conclusions} contains the discussion and conclusions.

\section{Methods}
\label{sec:methods}

This section presents the collection and processing of the imaging data and the mathematical model of tumor growth. We then described the proposed high-dimensional Bayesian framework that enables inferring the spatial distribution of model parameters from imaging data of individual subjects, accounting for uncertainties in both measurement and modeling. 
Fig. \ref{fig:overview} provides an overview of the framework.

\subsection{Experimental data}

\subsubsection{MRI data acquisition}
\label{sec:mri}
All experimental procedures were approved by the appropriate Institutional Animal Care and Use Committee.
Imaging data of glioma growth in four female Wistar rats are considered in this study, with the imaging and surgical procedures detailed in \cite{hormuth2017}. Briefly, $10^5$ C6 glioma cells were injected into the neocortex region of the animals' brains. 
The rats are imaged on different days after tumor implantation as listed in Table \ref{tab:mri}.
All MR images were acquired over a 32mm$\times$32mm$\times$16mm field of view sampled with a 128$\times$128$\times$16 matrix. 
A mutual information-based rigid registration algorithm was applied to register all MR scans to the first scan of each animal. 
This study uses $T_2$-weighted MR images to define the simulation domain and the contrast-enhanced $T_1$-weighted MRI for tumor segmentation. 
Diffusion-weighted MRI (DW-MRI) is used to compute apparent diffusion coefficient (ADC) maps  \cite{padhani2009} for each animal, which is then used to estimate the spatial distribution of tumor cells \cite{hormuth2017} (see Panel I of Fig.  \ref{fig:overview}). 
The carrying capacity (i.e., the maximum number of tumor cells that can exist within a given section of tissue) in each voxel is used to normalize the measurements and evaluate the spatio-temporal distribution of tumor volume fraction
$d(\boldsymbol{x}, t)$.
\vspace{-0.1in}
\begin{table}[h]
\caption{
The imaging measurements time points for each animal.
}
\center
\vspace{-0.1in}
\begin{scriptsize}
\begin{tabular}{l|l}
\hline
Animals & days post tumor implantation \\ \hline
\textit{Rat 1}   & 10, 12, 14, 15, 16, 18, 20   \\
\textit{Rat 2}   & 10, 12, 14, 15, 16, 18     \\
\textit{Rat 3}   & 10, 12, 14, 15, 16, 19   \\
\textit{Rat 4}   & 12, 14, 15, 16, 19, 21   \\ \hline
\end{tabular}
\end{scriptsize}
\label{tab:mri}
\vspace{-0.1in}
\end{table}
%

\subsubsection{Automated tissue segmentation}
\label{sec:segmentation}

Brain tissue heterogeneity governs the tumor’s spread and morphological development for each subject \cite{Gritsenko2012, giesexs1996, swan2018}. 
However, due to the limited spatial resolution of the acquired MRI data, tissue segmentation to grey matter, white matter, and cerebrospinal fluid in small animal brains is difficult to perform directly from the scans. 
To account for tissue heterogeneity,
this study proposes an automated method for tissue segmentation of a specific rat's brain, using available digital brain atlases along with a nonrigid local image registration algorithm. 
The spatial resolution and the level of contrast in the MRI dataset make some of the interior features, such as the corpus callosum, identifiable in the $T_2$-weighted images in each animal. The proposed method exploits matching these internal brain features in the atlas and individual rat scans for grey and white matter segmentation.  
Here, we use high-resolution Waxholm Space atlas images ($T_2$-weighted MRI and anatomical labeled images) of the Sprague Dawley rat brain presented in \cite{papp2014waxholm} and made available through the INCF Software Center 
(\href{http:// software.incf.org/software/waxholm-space-atlas-of-the-sprague-dawley-rat-brain}{http:// software.incf.org/software/waxholm-space-atlas-of-the-sprague-dawley-rat-brain}). 
For segmentation of a specific rat’s brain, we leverage the DEMONS image registration algorithm
\cite{thirion1998} that allows for large deformation and can track the variations between the images \cite{zheng2016}. This algorithm approximates nonparametric registration by a diffusion process and uses numerical optimization to compute a displacement vector at each voxel based on intensity differences \cite{kroon2020}. 
The resulting displacement field is convolved with a Gaussian kernel and iteratively transforms the \textit{moving (current)} image to the \textit{static (reference)} image. The proposed atlas-based segmentation method consists of the following steps:
\begin{itemize}
\item \textit{Image down-sampling:} 
The high-resolution $T_2$-weighted and anatomical atlas images are down-sampled (from 78$\times$103$\times$55 matrix) to match the resolution of the cropped brain image of each rat  (41$\times$61$\times$16 matrix).

\item \textit{Deformation map estimation:} 
The DEMONS algorithm is applied to $T_2$-weighted MRI scans of the specific rat (static image) and atlas (moving image).
The required displacement at each voxel is computed to match the external brain surface, corpus callosum, and other internal features in the two images.

\item \textit{Transforming atlas anatomical image:} The displacement vectors is then used to deform the grey and white matter of the anatomical image of the atlas to estimate the grey and white matter region in the specific rat.
\end{itemize}
Fig. \ref{fig:overview} (II.a) demonstrates different steps to segment a specific rat brain into the gray and white matter on the two-dimensional (2D) central slices along the cranial to the caudal direction.
%
The atlas-based segmentation method is implemented in MATLAB R2021b and using the built-in DEMONS algorithm.

\subsection{Tumor growth model}

The tumor model employed in this work is the 
standard single-species reaction-diffusion partial differential equation (PDE) \cite{murray2001, harpold2007}, capturing tumor proliferation and infiltration,
\begin{eqnarray}\label{eq:forward}
\frac{\partial u}{\partial t} = 
\nabla \cdot \left( D \nabla {u} \right) + 
G \left( 1 - {u} \right) u
& \text{in} & \Omega \times (0, T], \nonumber \\
D \nabla {u} \cdot \mathbf{n} = 0
& \text{on} & \partial \Omega \times (0, T], \nonumber \\
u (\boldsymbol{x}, 0) = u_0
& \text{in} & \Omega .
\end{eqnarray}
Here,
$\Omega$ is the brain domain of a particular rat, segmented from a 2D slice of $T_2$-weighted MRI of each rat,  and
$u(\boldsymbol{x},t) \in [0, 1]$ is the model estimated tumor volume fraction at each spatial point $\boldsymbol{x}$ in the brain domain $\Omega$ and time $t$ [day] with $T$ being the final time.
Parameter ${D} (\boldsymbol{x}) $ accounts for the tumor spread in the host tissue due to invasion and cell migration, and
$G(\boldsymbol{x})$ denotes the proliferation rate of the tumor, the rate at which tumor cells grow and divide.
The domain boundary $\partial \Omega$ with outward unit normal $\mathbf{n}$ corresponds to the inner surface of the skull and is not infiltrated by the tumor cells. Finally,
$u_0(\boldsymbol{x})$ is the initial tumor volume fraction.
The model parameters $\boldsymbol{\theta} = \left( \log D(\boldsymbol{x}), \log G(\boldsymbol{x}) \right)$ are considered unknown, subject-specific, and spatially varying to account for the brain tissue and tumor heterogeneities. 
Here, the model parameters represent the logarithm of the tumor diffusion and proliferation coefficients, as this ensures the positivity of these estimated quantities in the subsequent Bayesian inversion.
The solution of the PDE \eqref{eq:forward} is approximated \textit{via} the nonlinear finite element method, using FEniCS \cite{fenics}.

\subsection{High-dimensional Bayesian calibration of tumor model}
\label{sec:bayes}

We use Bayesian inference to calibrate the model parameters $\boldsymbol{\theta}$ using data $\mathbf{d}$ with quantified uncertainties in measurements and modeling errors.
Bayes' theorem \cite{jaynes2003} is written as,
\begin{equation}\label{eq:bayes}
\pi_{\rm post}(\boldsymbol{\theta}|\mathbf{d}) \propto {\pi_{\rm like}(\mathbf{d}|\boldsymbol{\theta}) \pi_{\rm pr}(\boldsymbol{\theta})}
\end{equation}
where $\pi_{\rm pr}(\boldsymbol{\theta})$ is the \textit{prior} probability distribution function (PDF) incorporating any prior information about the values and spatial distributions of the parameters.
The \textit{likelihood} PDF $\pi_{\rm like}(\mathbf{d}|\boldsymbol{\theta})$ in \eqref{eq:bayes} 
is derived from the noise model describing uncertainty in the 
data and modeling error.
The \textit{posterior} PDF $\pi_{\rm post}(\boldsymbol{\theta}|\mathbf{d})$ is the updated prior of the parameters using the data $\mathbf{d}$.
We note that since the model parameters are spatial fields, after discretization,
$\boldsymbol{\theta} \in \mathbb{R}^N$ with $N$ being the number of finite element nodes discretizing the brain domain.
This gives rise \eqref{eq:bayes} to a high-dimensional Bayesian inference problem.

\subsubsection{Constructing the subject-specific prior}

Accurate inference of the model parameter of glioma growth in a specific subject requires incorporating all the relevant information to construct meaningful prior PDFs. 
Due to the nerve fiber bundles in white matter, it is hypothesized that the spread of the tumor (represented by $D$ in the model) is faster in this region than in the grey matter \cite{Gritsenko2012, swan2018}. 
Moreover, the higher blood volume in the grey matter \cite{lee2006} may result in a higher proliferation rate of the tumor (represented by $G$ in the model) in the grey matter.
The common approach in the literature is assigning different model parameter values to grey and white matter.
In this work, we assume model parameters to be spatially varying fields with different priors in the gray and white matter regions to be updated by each rat's imaging data. 
In particular, we consider the prior to be a Gaussian random field with the Mat\'ern covariance kernel \cite{lindgren2011}, that is $\boldsymbol{\theta} \sim \mathcal{N}( \bar{\boldsymbol{\theta}}, \mathcal{C}_{\rm pr} )$ where $ \bar{\boldsymbol{\theta}}$ is the prior mean  and $\mathcal{C}_{\rm pr}$ is the covariance operator.
Following \cite{bui2013}, we define $\mathcal{C}_{\rm pr} = \mathcal{A}^{-2}$ as the squared inverse of the differential operator $\mathcal{A}$ given by 
\begin{equation}\label{eq:matern}
\mathcal{A} \boldsymbol{\theta} = \left\{
\begin{array}{ll}
-\gamma {\Delta} \boldsymbol{\theta} + \delta \boldsymbol{\theta}  & \text{in}   \quad \Omega,\\
\beta\boldsymbol{\theta} + \nabla\boldsymbol{\theta} \cdot \mathbf{n}
& \text{on} \quad \partial \Omega,
\end{array}
\right.
\end{equation}
where $\nabla$ and ${\Delta}$ are the  gradient and Laplace operators, respectively.
Samples from the prior can then be drawn by numerically solving the stochastic PDE, $\mathcal{A}(\boldsymbol{\theta} - \bar{\boldsymbol{\theta}}) = W$, where $ W$ denotes the spatial Gaussian white noise with unit variance.
The spatially varying parameters $\delta$ and $\gamma$ control 
the prior variance $\sigma^2$, 
the correlation length $\rho$ between two spatial points of the parameters, and
the optimal Robin coefficient $\beta$ to minimize boundary artifacts \cite{VillaPetraGhattas18}, such that
\begin{equation}
\delta = \frac{\sqrt{2}}{\sigma \rho \sqrt{\pi}} 
\quad, \quad
\gamma = \frac{1}{4\sqrt{2\pi}} \frac{\rho}{\sigma}
\quad, \quad
\beta = \frac{\sqrt{\delta\gamma}}{1.42}.
\end{equation}
The subject-specific prior of the tumor model parameters are constructed by considering different prior means $\bar{\boldsymbol{\theta}}$ and correlation lengths $\rho$ 
of the parameters in the grey and white matter to represent different diffusion and proliferation mechanisms in these regions. The proposed atlas-based brain segmentation provides the most probable region of the white and grey matter in individual rats. However, uncertainties are associated with the identified grey and white matter interface due to the errors in image processing, the digital atlas, and possible deformation of the white matter during tumor growth. To account for these uncertainties, a small spatial correlation length is imposed at the region between the grey and white matter representing the uncorrelated parameters across the interface.

\subsubsection{Likelihood}
We define the noise model as,
$d(\boldsymbol{x}, t_i)  = u(\boldsymbol{x}, t_i)  + \eta_{T}$, 
where 
$d(\mathbf{x},t_i)$ is the spatial distribution of tumor volume fraction from MRI and acquired at times $t_i$, $u(\mathbf{x},t_i)$ is the corresponding solutions of the forward model \eqref{eq:forward}, and
$\eta_T$ is the total error that accounts for  modeling and measurement uncertainties.
Assuming no systematic bias in the computational model and data acquisition, we assume $\eta_{T} \sim \mathcal{N}(0, \sigma^2_{\rm noise}\mathbf{I})$, where $\sigma^2_{\rm noise}$ is the unknown noise variance \cite{kennedy2001}.
The form of the likelihood is then given by,
\begin{equation}\label{eq:like}
\pi_{\rm like}(\mathbf{d}|\boldsymbol{\theta}) 
\propto \exp\left\{ \sum_i \frac{-1}{\sigma^2_{\rm noise}}\int_{\Omega}( u(\boldsymbol{x}, t_i) - d(\boldsymbol{x}, t_i) )^2 \,d\boldsymbol{x} \right\}.
\end{equation}
%

\subsubsection{Scalable solution of Bayesian calibration}

The dependency of the number of parameters of the tumor model on the finite element discretization leads to a high computational cost of the Bayesian inference using standard solutions. 
To overcome this challenge, we make use of a scalable solution algorithm relying on a Laplace approximation of the posterior \cite{issac2015},
\begin{equation}\label{eq:LA}
\pi^L_{\rm post}(\boldsymbol{\theta}|\mathbf{d}) \propto
\exp \left\{- \frac{1}{2} (\boldsymbol{\theta} - \boldsymbol{\theta}_{\rm MAP})^T \boldsymbol{\Gamma}^{-1}_{\rm post} (\boldsymbol{\theta} - \boldsymbol{\theta}_{\rm MAP}) \right\},
\end{equation}
that reduces computing the posterior PDF to the evaluation of the 
maximum a posterior (MAP) estimate $\boldsymbol{\theta}_{\rm MAP}$,
and the posterior covariance $\boldsymbol{\Gamma}_{\rm post}$ at each spatial point.
%
We compute the $\boldsymbol{\theta}_{\rm MAP}$ by minimizing the negative log-posterior,
\begin{equation}\label{eq:optimization}
\boldsymbol{\theta}_{\rm MAP} := 
\underset{\boldsymbol{\theta}\in \mathbb{R}^N}{\text{argmin}} \; \{-\log(\pi_{\rm post}(\boldsymbol{\theta}|\mathbf{d})) \},
\end{equation}
using an adjoint-based globalized Newton Conjugate Gradient algorithm \cite{petra2012inexact}.
%
The posterior covariance is given by
\begin{equation}\label{eq:post_cov}	
\boldsymbol{\Gamma}_{\rm post}  = \left(\mathbf{H}(\boldsymbol{\theta}_{\rm MAP}) +  \boldsymbol{\Gamma}_{\rm pr}^{-1}\right)^{-1}
\end{equation}
where
$\boldsymbol{\Gamma}_{pr}$ is the prior covariance matrix (i.e., the discrete counterpart of $\mathcal{C}_{\rm pr}$), and
$\mathbf{H}(\boldsymbol{\theta}_{\rm MAP})$ is the Hessian matrix of the negative log-likelihood evaluated at the MAP.
To make computation involving $\boldsymbol{\Gamma}_{\rm post}$ feasible we adopt the low-rank based approximation in   
\cite{issac2015}. 
First, using randomized algorithms \cite{halko2011} we estimate the $r$ dominant eigenpairs  $\{\lambda_i\}_{i=1}^r$, $\{\mathbf{v}_i\}_{i=1}^r$
 of the generalized eigenvalue problem
$
\mathbf{H} \mathbf{v}_i = \lambda_i \boldsymbol{\Gamma}_{\rm pr}^{-1} \mathbf{v}_i.
$
Then we apply the Sherman-Morrison-Woodbury identity to yield the approximation 
$\boldsymbol{\Gamma}_{\rm post}  \approx \boldsymbol{\Gamma}_{\rm pr} - \boldsymbol{V} \boldsymbol{D} \boldsymbol{V}$, where the columns of $\boldsymbol{V}$ collect the generalized eigenvectors $\mathbf{v}_i$ and $\boldsymbol{D}$ is a diagonal matrix with entries $\lambda_i(1+\lambda_i)^{-1}$.
Since $r$ is much smaller and independent of the number of parameters $N$, this approximation results in a scalable solution of the high-dimensional Bayesian inversion (see \cite{issac2015} for more details).
%
The algorithm for the inference of the time-dependent nonlinear tumor model \eqref{eq:forward} is implemented in the hIPPYlib \cite{VillaPetraGhattas21, VillaPetraGhattas18}.

\subsubsection{Hyper-parameters}
The proposed high-dimensional Bayesian framework consists of a set of hyper-parameters $\boldsymbol{\vartheta}$ $=$ $( \bar{\boldsymbol{\theta}}_{wm}$, $\bar{\boldsymbol{\theta}}_{gm}$, $\sigma_{wm}$, $\sigma_{gm}$, $\rho_{wm}$, $\rho_{gm}$, $\rho_{int}$, $\sigma_{noise})$, where the subscripts $gm$, $wm$, and $int$ indicate the parameters corresponding to the grey matter, white matter, and their interface, respectively. 
The parameter correlation lengths and the prior means and variances are specific to the Wistar rats, while noise variance depends on the model and medical imaging techniques.
The approach for estimating these hyper-parameters is described in Section \ref{sec:hyper}.

\section{Results}
\label{sec:results}

We demonstrate the process of model calibration from 
each rat's MRI data using the 
high-dimensional Bayesian framework described in Section \ref{sec:bayes}.
All model calibrations were conducted in 2D using the center slice of the tumor along the cranial to caudal direction. 
For finite element solutions of the tumor growth model \eqref{eq:forward}, the brain domains of the animals are discretized with triangular elements.
%
Each rat's tumor volume fraction at the first imaging time is used to initialize the tumor model $u_0$. 
Consequently, the imaging data at multiple time points are used to calibrate the parameter fields (training set) using the Bayesian inference and the subject-specific prior. The model’s predictions are then tested by comparing with the imaging data in later time points, which were excluded from the calibration (unseen testing set). 
\vspace{-0.1in}
\begin{figure}[!h]
\centerline{\includegraphics[width=\columnwidth]{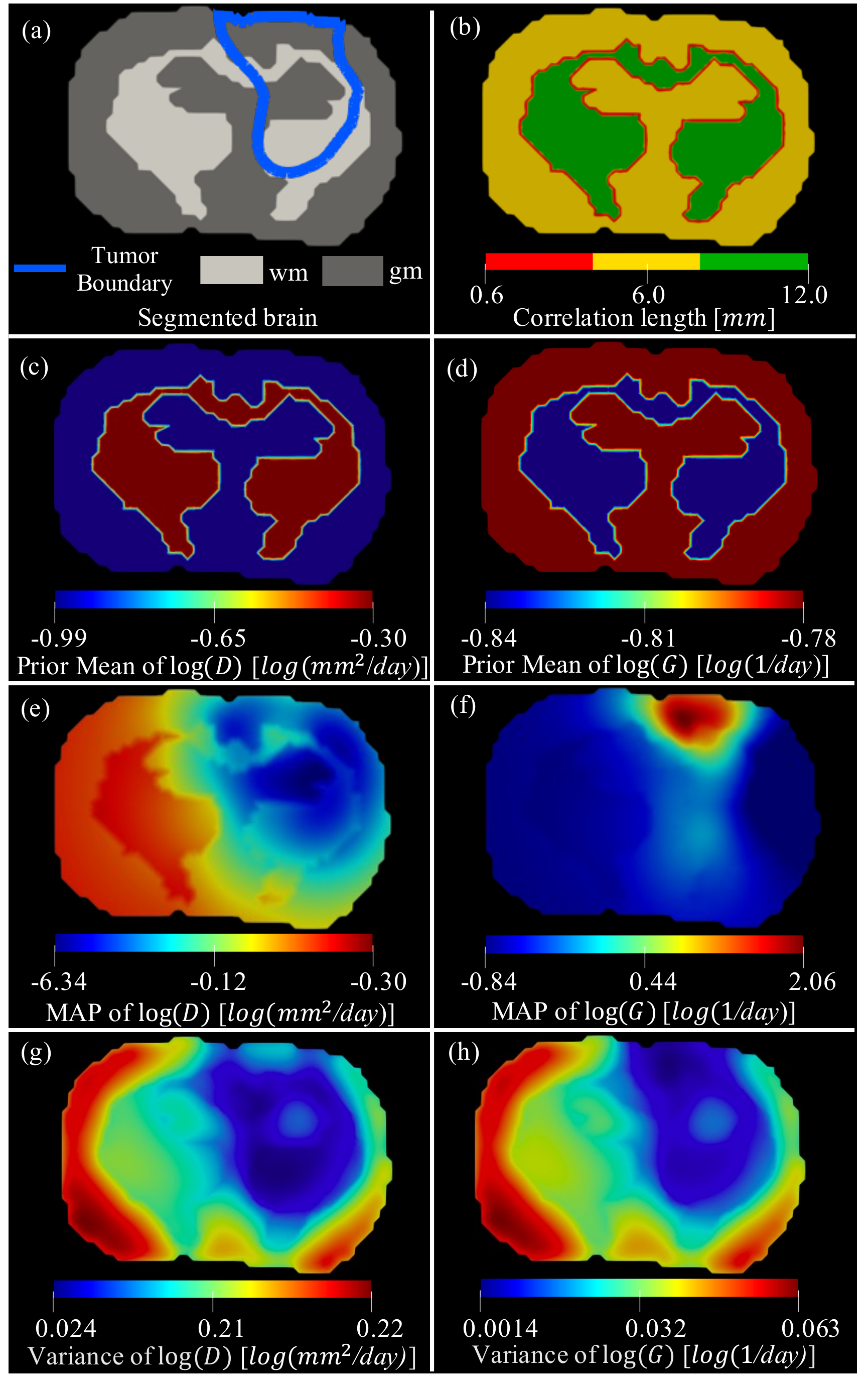}}
\vspace{-0.1in}
\caption{
	Results of subject-specific prior construction and Bayesian calibration of \textit{Rat 3} using the training imaging data points on days 12, 14, 15, and 16. 
	Panel (a) displays the grey and white matter regions ($gm$ and $wm$) obtained from the atlas-based image segmentation method along with the computed tumor boundary on day 16. 
	Panel (b) shows the parameter correlation length assigned to each region, while 
	panels (c) and (d) present the prior means of the diffusion $\text{\textit{D}}$ and proliferation rate $\text{\textit{G}}$.
	Panels (e)-(h) show MAP and variance of the posteriors, obtained \textit{via} Bayesian inference, indicating that the model parameters are learned from data in the surrounding region of the growing tumor.
}
\label{fig:prior}
\vspace{-0.15in}
\end{figure}

\vspace{-0.0in}
\subsection{Hyper-parameters estimation}
\label{sec:hyper}

The estimated hyper-parameters for Bayesian calibration of the tumor model for all animals are listed in Table \ref{tab:prior}.
The prior means and variances of the model parameters in gray and white matter are obtained from the averaged posterior distributions from the previous studies on similar Wistar rats \cite{lima2017selection}, along with considering the $D_{wm}/D_{gm} = 2$ as estimated in \cite{hormuth2015predicting} and $G_{wm}/G_{gm} = 0.94$ based on the blood volume of the brain reported in \cite{thalman2019brain}.
Figure \ref{fig:prior} (a), (c), and (d) show the segmented brain of \textit{Rat 3} to grey and white matter using the atlas-based method and the assigned prior means.
\begin{table}[h]
\vspace{-0.1in}
\caption{
Estimated hyper-parameters $\boldsymbol{\vartheta}$ of the Bayesian framework.
}
\vspace{-0.05in}
\setlength{\tabcolsep}{3pt}
\centering
\begin{tabular}{cccccccc}
\hline
\multicolumn{8}{c}{Prior mean and variance of parameters}                                                                                                                                                                                                                                                                                                          \\ \hline
\multicolumn{2}{c|}{\begin{tabular}[c]{@{}c@{}}log($D_{gm}$)\\ log($mm^2/day$)\end{tabular}} & \multicolumn{2}{c|}{\begin{tabular}[c]{@{}c@{}}log($D_{wm}$)\\ log($mm^2/day$)\end{tabular}} & \multicolumn{2}{c|}{\begin{tabular}[c]{@{}c@{}}log($G_{gm}$)\\ log($1/day$)\end{tabular}} & \multicolumn{2}{c}{\begin{tabular}[c]{@{}c@{}}log($G_{wm}$)\\ log($1/day$)\end{tabular}} \\
Mean                                & \multicolumn{1}{c|}{Variance}                             & Mean                                & \multicolumn{1}{c|}{Variance}                             & Mean                             & \multicolumn{1}{c|}{Variance}                          & Mean                                       & Variance                                    \\
-0.9937                             & \multicolumn{1}{c|}{0.2336}                               & -0.3006                             & \multicolumn{1}{c|}{0.2336}                               & -0.7800                          & \multicolumn{1}{c|}{0.0682}                            & -0.8419                                    & 0.0682                                      \\ \hline
\multicolumn{8}{c}{Spatial correlation lengths of parameters $\boldsymbol{\theta}$ and noise variance}                                                                                                                                                                                                                                                                                                                                               \\ \hline
\multicolumn{2}{c|}{$\rho_{gm}$ ($mm$)}                                                         & \multicolumn{2}{c|}{$\rho_{wm}$ ($mm$)}                                                         & \multicolumn{2}{c|}{$\rho_{int}$ ($mm$)}                                                  & \multicolumn{2}{c}{$\sigma_{noise}^2$}                                                   \\
\multicolumn{2}{c|}{6.0}                                                                        & \multicolumn{2}{c|}{12.0}                                                                       & \multicolumn{2}{c|}{0.6}                                                                  & \multicolumn{2}{c}{3.9E-3}                                                               \\ \hline
\end{tabular}
\label{tab:prior}
\vspace{-0.1in}
\end{table}
%

The correlation lengths and noise variance are estimated \textit{via} a cross-validation approach and using the \textit{Rat 1} and \textit{Rat 2} data. The considered initial ranges are $[2 \text{mm},10 \text{mm}]$ for $\rho_{gm}$ based on the brain size, $[0.5, 1]$ for $k = \rho_{gm}/\rho_{wm}$ due to the existence of nerve fibers and thus stronger spatial correlations in the white matter, and $[0.015, 0.5]$ for $\sigma_{noise}$. The grid search optimization is then performed such that for each combination of the hyper-parameters, model calibration is performed using the DW-MRI data in the training sets (days 12 - 18 for \textit{Rat 1} and days 12 - 16 for \textit{Rat 2}). 
\begin{figure*}[h]
    \centering
 \includegraphics[width=0.90\textwidth]{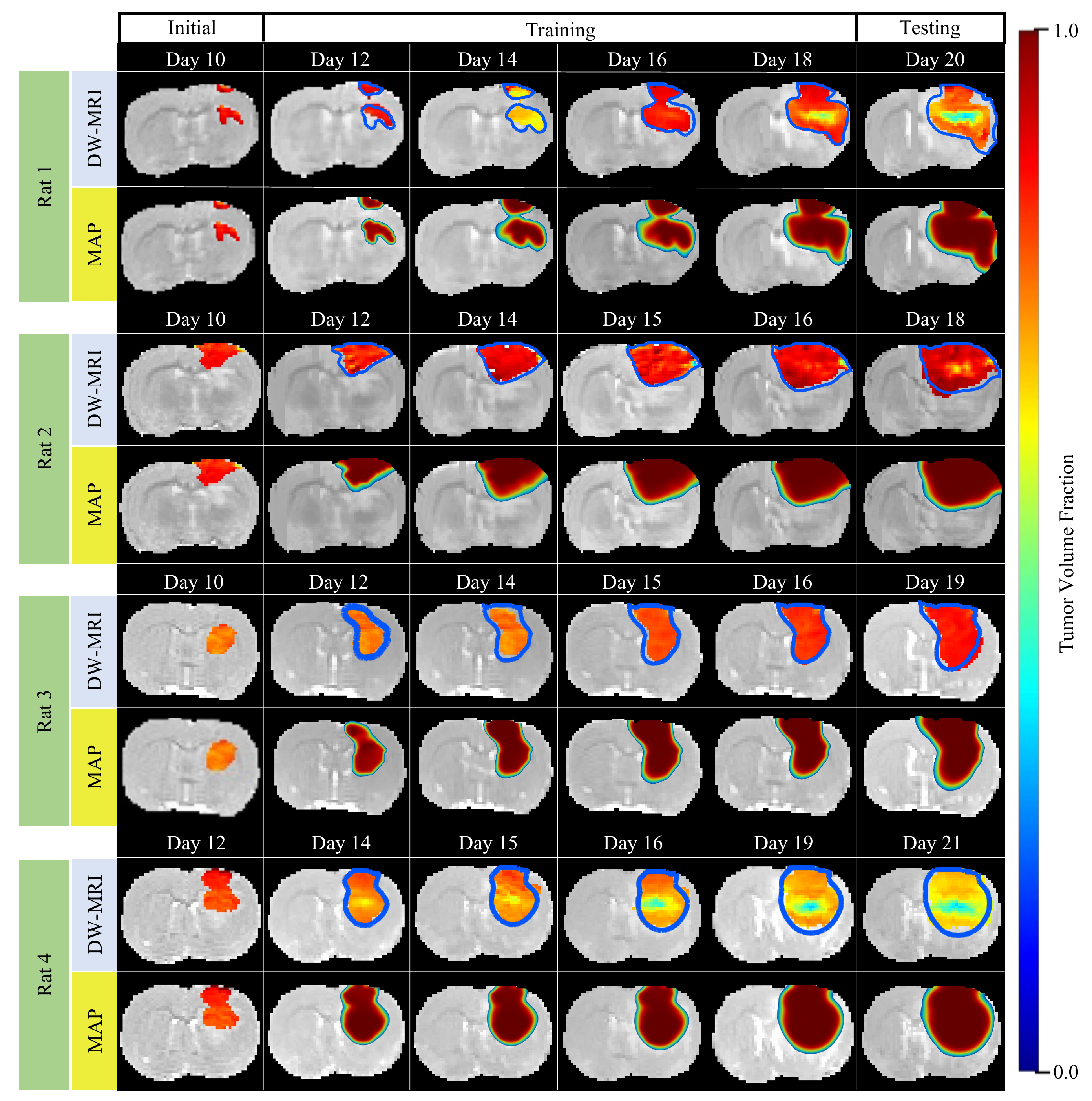}
  \vspace{-0.1in}
 \caption{
 Experimentally measured and model predicted tumor volume fraction using the Bayesian framework in four rats at the indicated time points after tumor implantation. For each animal, the measured tumor volume fractions from the DW-MRI data (top row) and the corresponding computational model prediction using the MAP values of the parameters (bottom row) are shown.
The blue outlines indicate the most probable model predicted tumor boundaries.
The data in the training set are used in the Bayesian calibration process (\textit{Rat 1} includes another training data at day 15 not shown in this figure). 
 The testing set is the unseen time point excluded from the calibration to assess the model's predictive ability to capture future tumor morphology. 
 The Dice-overlap between the predicted tumor shape with the imaging data at the testing time points are
0.91$\pm$0.01,
 0.92$\pm$0.01,
 0.90$\pm$0.01, and
 0.92$\pm$0.01 for
 \textit{Rat 1} to \textit{Rat 4}, respectively.
 }
 \label{fig:state}
 \vspace{-0.2in}
\end{figure*}
The error in normalized tumor area (NTA) and Dice similarity coefficient between the model prediction and MRI data at the testing sets (day 20 for \textit{Rat 1} and day 18 for \textit{Rat 2}) are taken into account as the figures of merit. 
We define a tumor indicator function with a value of one in the regions with non-zero tumor volume fractions and zero otherwise. The NTA error is then obtained as the distance between the tumor indicator in the data and the model divided by the brain area.
Dice ranges from 0 (no spatial overlap) to 1 (complete overlap) and characterizes the agreement between computed and measured tumor shapes.
The solutions to the multi-objective optimization are then obtained using Pareto sets to maximize Dice and minimize NTA error in both rats, resulting in the optimal $\rho_{wm}$, $\rho_{gm}$, $\sigma_{noise}$ listed in Table \ref{tab:prior}. 
Finally, a short correlation length of 0.6mm is imposed near the interface between the grey and white matter, as shown in Fig. \ref{fig:prior}(b). The width of this region accounts for the uncertainty in identifying the grey and white matter interface from the segmentation method and its possible movement during tumor growth.
The values of hyperparameters reported in Table \ref{tab:prior} can be used for predictive modeling of tumors in other Wistar rats using the same tumor model and imaging protocol.
%

\vspace{-0.1in}
\subsection{Bayesian calibration of the tumor model}
Figure \ref{fig:prior}(e-h) shows the posteriors' MAP and point-wise variance of the model parameters for \textit{Rat 3} determined from the corresponding training data sets; i.e., measured tumor volume fractions on days 12 - 16 after tumor implantation.
Comparing the priors and posteriors of diffusion and proliferation rates in Fig. \ref{fig:prior} indicates that the model parameters are learned from data, leading to a smaller variance around the tumor region.
Figure \ref{fig:state} compares the tumor volume fractions estimated using DW-MRI with that predicted by the computational model evaluated at the parameters' MAP. 
The subject-specific model calibration depicted in this figure indicates that the spatially varying parameters and correlation of the parameters built into the priors \eqref{eq:matern} of the Bayesian framework enables the model to capture the tumor spread pattern in all rats. 
Additionally, the tumor model, constrained by the animal anatomy and the training data, can accurately predict the irregular tumor morphologies observed in the testing data (Dice $>$ 0.9 in all rats) along with the most probable tumor infiltration into the surrounding brain tissue.
We note that the uncertainties in the MRI data of the testing set, due to imaging and registration to the first scan, contribute to the reported model prediction errors. 
A particular source of data uncertainty, for example, is the tumor shrinkage in \textit{Rat 1} between days 18 and 20, a phenomenon that violates the tumor growth mechanism built into the model.

\vspace{-0.0in}
\subsection{Uncertainty in tumor growth predictions}

\begin{figure}[!t]
	\centerline{\includegraphics[trim=0 0.0in 0.2in 0.1in, clip, width=\columnwidth]{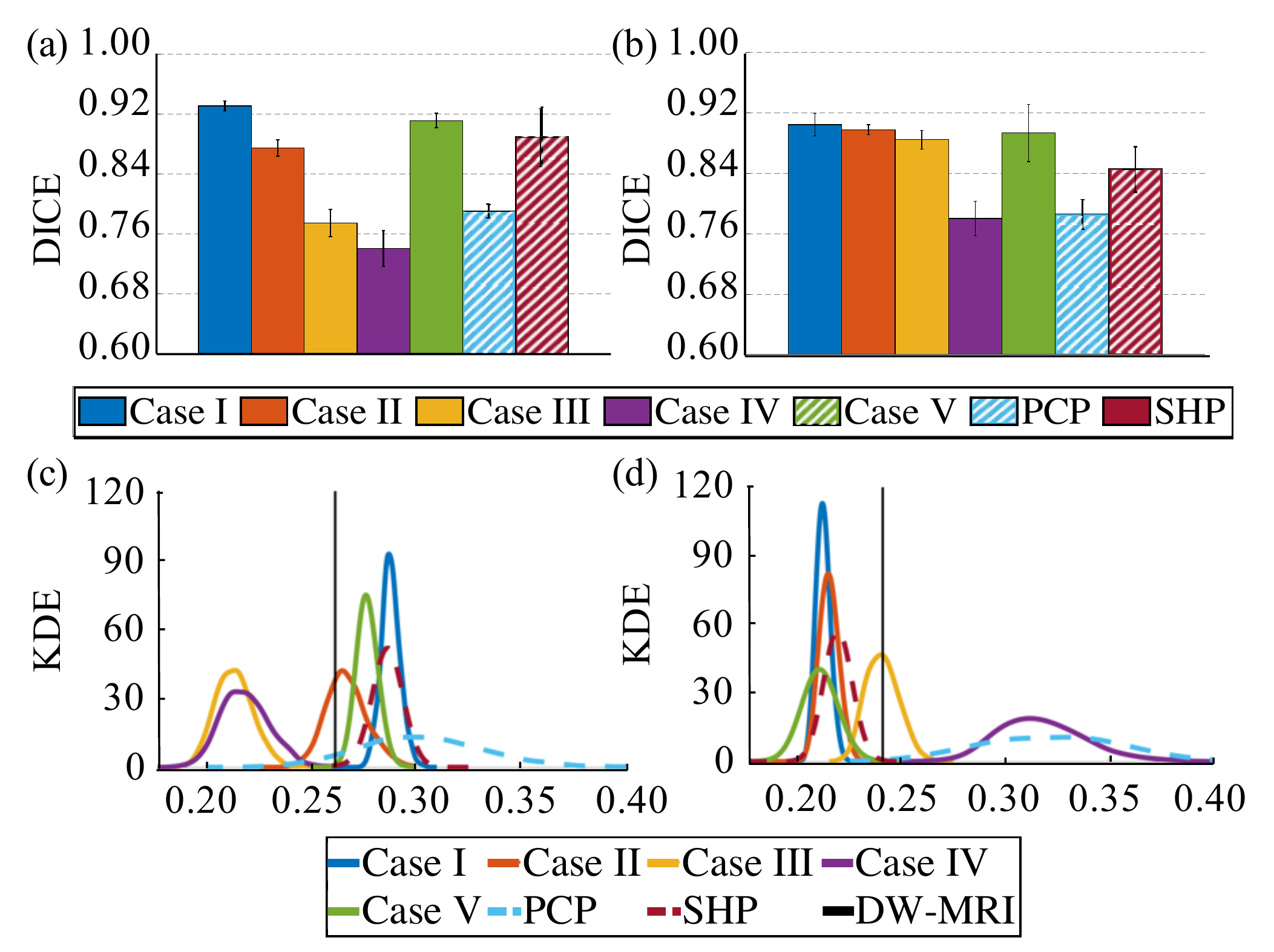}}
	\caption{
		Quantitative assessment of the predictive accuracy at the testing time points in \textit{Rat 1} at day 20 (left column) and \textit{Rat 3} at day 18 (right column) for the different calibration scenarios in Table \ref{tab:prediction}. 
		Panels (a) and (b) present the Dice overlap between measured data and model prediction.  Panels (c) and (d) compare the measured NTA (straight lines) and kernel density estimate (KDE) of the NTA predicted using the proposed Bayesian calibration framework. 
		The results are from 1000 samples of the parameter posteriors, and the error bars represent 95\% confidence intervals. In each rat, statistically significant (\textit{p} $<$ 4.5E-3) differences in Dice are observed between all the scenarios.
	}
	\label{fig:kde}
\end{figure}

The animal-specific predictions of tumor morphology presented in Fig. \ref{fig:state} correspond to deterministic model calibrations and do not address the reliability of model prediction in the presence of imaging and modeling uncertainties. Conversely, the Bayesian framework enables propagating parametric uncertainty through the model to obtain robust posterior predictions about the tumor morphology.
Table \ref{tab:prediction} and Fig. \ref{fig:kde} compare the predictive ability of the tumor model for different cases in \textit{Rat 1} and \textit{Rat 3} (recall that data of \textit{Rat 1} is used for determining the hyper-parameters in Section \ref{sec:hyper}).
In all cases, the first imaging time point defines the initial tumor volume fraction $u_0$ in the computational model for each rat. The last time points (testing sets) are used to assess model prediction by computing the Dice and relative error in NTA compared to the imaging data.

\begin{table}[]
\caption{
The relative error between the measured and computed tumor for different calibration scenarios.
}
\vspace{-0.1in}
\center
\begin{scriptsize}
\begin{tabular}{ccccccc}
\hline
\multirow{2}{*}{Animals}                 & \multirow{2}{*}{Cases} & \multicolumn{3}{c}{Days post tumor implantation}                                        & \multicolumn{2}{c}{Prediction Error} \\ \cline{3-7} 
                                         &                        & Initial                 & Training                            & Testing                 & NTA               & Dice             \\ \hline
\multicolumn{1}{c|}{\multirow{7}{*}{Rat 1}} & I                      & \multicolumn{1}{c|}{10} & \multicolumn{1}{c|}{12,14,15,16,18} & \multicolumn{1}{c|}{20} & .10$\pm$.02     & .91$\pm$.01    \\
\multicolumn{1}{c|}{}                    & II                     & \multicolumn{1}{c|}{10} & \multicolumn{1}{c|}{12,14,15,16}    & \multicolumn{1}{c|}{20} & .03$\pm$.03     & .88$\pm$.01    \\
\multicolumn{1}{c|}{}                    & III                    & \multicolumn{1}{c|}{10} & \multicolumn{1}{c|}{12,14,15}       & \multicolumn{1}{c|}{20} & .18$\pm$.03     & .77$\pm$.02    \\
\multicolumn{1}{c|}{}                    & IV                     & \multicolumn{1}{c|}{10} & \multicolumn{1}{c|}{12,14}          & \multicolumn{1}{c|}{20} & .16$\pm$.04     & .74$\pm$.02    \\
\multicolumn{1}{c|}{}                    & V                      & \multicolumn{1}{c|}{10} & \multicolumn{1}{c|}{16,18}          & \multicolumn{1}{c|}{20} & .07$\pm$.08     & .91$\pm$.06    \\ \cline{2-7} 
\multicolumn{1}{c|}{}                    & SHP                    & \multicolumn{1}{c|}{10} & \multicolumn{1}{c|}{12,14,15,16,18} & \multicolumn{1}{c|}{20} & .10$\pm$.02     & .89$\pm$.04    \\ 
\multicolumn{1}{c|}{}                    & PCP                    & \multicolumn{1}{c|}{10} & \multicolumn{1}{c|}{12,14,15,16,18} & \multicolumn{1}{c|}{20} & .17$\pm$.10     & .79$\pm$.01    \\
\hline
\multicolumn{1}{c|}{\multirow{7}{*}{Rat 3}} & I                      & \multicolumn{1}{c|}{10} & \multicolumn{1}{c|}{12,14,15,16}    & \multicolumn{1}{c|}{18} & .12$\pm$.04     & .90$\pm$.01    \\
\multicolumn{1}{c|}{}                    & II                     & \multicolumn{1}{c|}{10} & \multicolumn{1}{c|}{12,14,15}       & \multicolumn{1}{c|}{18} & .11$\pm$.02     & .90$\pm$.01    \\
\multicolumn{1}{c|}{}                    & III                    & \multicolumn{1}{c|}{10} & \multicolumn{1}{c|}{12,14}          & \multicolumn{1}{c|}{18} & .03$\pm$.02     & .89$\pm$.01    \\
\multicolumn{1}{c|}{}                    & IV                     & \multicolumn{1}{c|}{10} & \multicolumn{1}{c|}{12}             & \multicolumn{1}{c|}{18} & .32$\pm$.08     & .79$\pm$.02    \\
\multicolumn{1}{c|}{}                    & V                      & \multicolumn{1}{c|}{10} & \multicolumn{1}{c|}{16}             & \multicolumn{1}{c|}{18} & .19$\pm$.19     & .89$\pm$.10    \\ \cline{2-7} 
\multicolumn{1}{c|}{}                    & SHP                    & \multicolumn{1}{c|}{10} & \multicolumn{1}{c|}{12,14,15,16}    & \multicolumn{1}{c|}{18} & .09$\pm$.01      & .84$\pm$.03    \\ 
\multicolumn{1}{c|}{}                    & PCP                    & \multicolumn{1}{c|}{10} & \multicolumn{1}{c|}{12,14,15,16}    & \multicolumn{1}{c|}{18} & .33$\pm$.14      & .79$\pm$.02    \\
\hline
\end{tabular}\end{scriptsize}
\label{tab:prediction}
\vspace{-0.25in}
\end{table}

In the \textit{Case I} to \textit{Case IV}, one imaging time point is sequentially discarded from the end of the training set. 
The results of these cases indicate lower accuracy and confidence in tumor shape predictions in both rats by increasing the time gap between the last training data time point and subject-specific prediction at the testing times (a 12.22\% decrease in the means and a 100\% increase in standard deviations of Dice in \textit{Case I} compared to \textit{Case IV} of \textit{Rat 3}). Similarly, the variances of NTA monotonically increase by the successive exclusion of training data points, leading to the sharper kernel density estimations of the \textit{Case I} in Fig. \ref{fig:kde} (c) and (d). 
However, the probability distributions of NTA do not precisely represent the prediction accuracy of tumor spatial patterns in different cases. 

%

\begin{figure}[!t]
\centerline{\includegraphics[trim=0 0 0 0.1in,clip, width=\columnwidth]{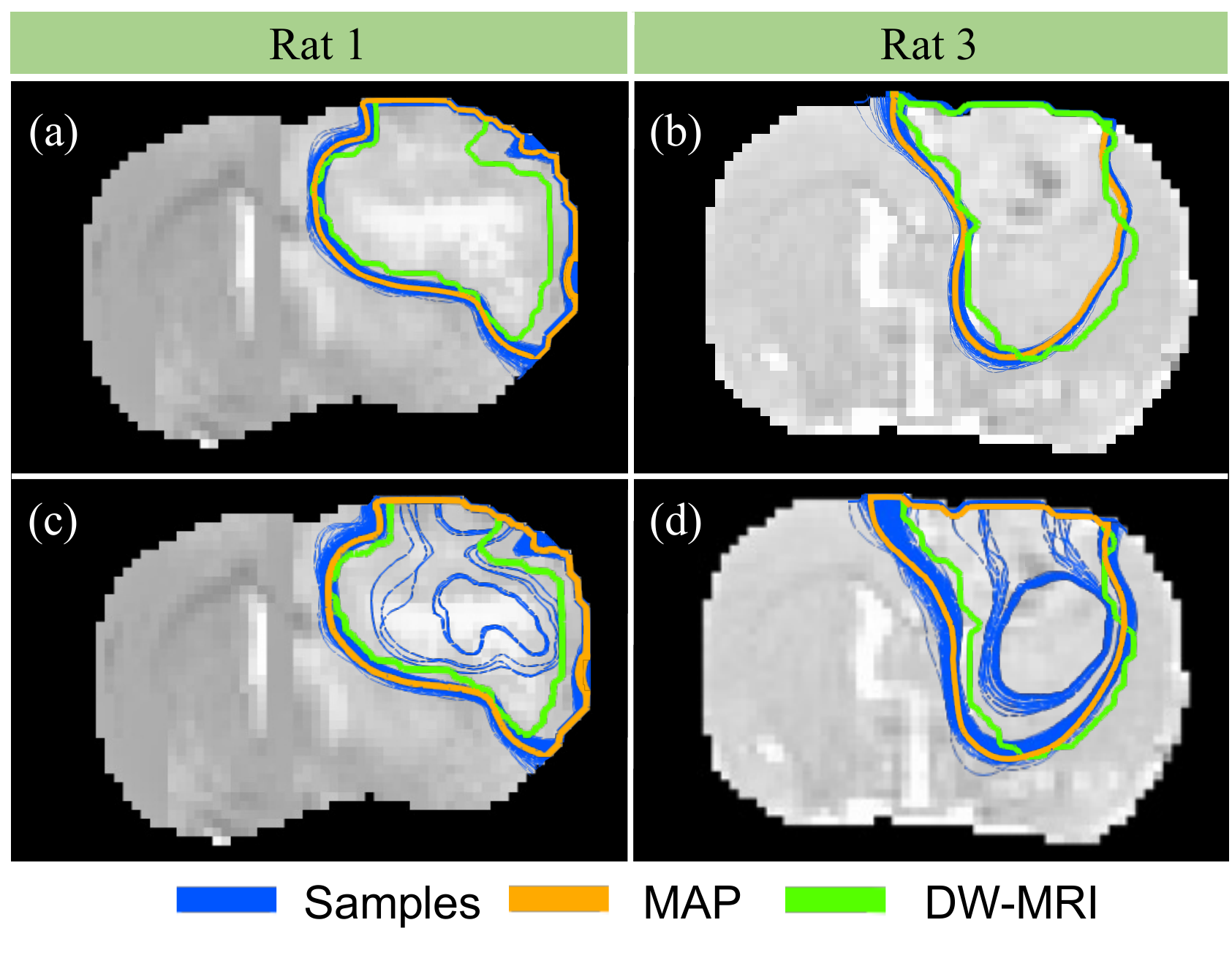}}
\vspace{-0.1in}
\caption{
Panels (a) and (c) compare tumor boundaries for \textit{Rat 1} obtained from the MRI data (yellow lines) and model prediction at the testing time point of day 20 using parameters' MAP (green lines) and 250 samples from parameter posteriors (blue lines).  
Panels (b) and (d) present similar results for \textit{Rat 3} at day 18.
The top row shows the model prediction results obtained from the \textit{Case I} calibration, and the bottom row shows the results of \textit{Case V}, according to Table \ref{tab:prediction}.
Lower confidence in the computational predictions of the tumor morphologies in \textit{Case V} is recognized from the broader spread of the possible tumor boundary throughout the domain and as a result of the smaller number of training data at earlier time points compared to \textit{Case I}.
}
\label{fig:samples}
\vspace{-0.25in}
\end{figure}

In the \textit{Case V}, the tumor model is calibrated using the imaging data only on days 16 and 18 for \textit{Rat 1} and day 16 for \textit{Rat 3}. This case aims to assess the predictive power of the proposed framework when the tumor is only diagnosed at a late stage. Remarkably, in both rats, the predicted NTA and Dice values in \textit{Case V} are close to the ones in the \textit{Case I} where all the available imaging data between the initial and testing time points are used in calibration (the relative error of $1.1\%$ in the Dice between the \textit{Case I} and \textit{Case V} in \textit{Rat 3}). 
Such predictive power of the model using fewer data is attributed to the parameters' spatial correlation that encodes surrounding tissue information that drives heterogeneous tumor expansion into the calibrated model. However, the standard deviation of Dice values in \textit{Case V} are noticeably (6 times in \textit{Rat 1}) higher than \textit{Case I} as shown in Fig. \ref{fig:kde} (a) and (b).
Figure \ref{fig:samples} compares tumor boundary derived from the MRI scan at the testing time points (day 20 for \textit{Rat 1} and day 18 for \textit{Rat 3}) with the corresponding most probable tumor outlines computed using the parameters MAP and 250 samples drawn from the parameter posteriors in \textit{Case I} and \textit{Case V}. 
The computed tumor outlines are considered with the cut-off volume fraction of 0.5 \cite{le2016mri}, excluding the tumor infiltration, to be comparable with the visible tumor in MRI.
Consistent with Fig. \ref{fig:kde} and Table \ref{tab:prediction}, in each rat, the predicted tumor boundaries using MAPs are very close to those estimated from MRI data in both cases. In the \textit{Case I}, all the posterior prediction samples of the tumor boundary are near the tumor outlines obtained from the MAPs, with an average boundary margin of 0.58 mm.
However, samples of tumor boundaries predicted by the model are more spread throughout the domain, indicating lower confidence in the prediction of the tumor morphology in \textit{Case V}.
Overall results of four rats indicate that the proposed Bayesian framework enables accurate prediction of tumor morphologies within approximately four days after the last training data point. However, the reliability of such computational prediction strongly relies on the number of training imaging data at earlier time points.
%

\subsection{Comparison with other calibration methods}

This section compares the subject-specific tumor model predictions from the proposed high-dimensional Bayesian framework with two possibly more straightforward calibration methods. 
The first approach is \textit{spatially homogeneous prior} (SHP), 
in which we considered the same prior means and correlation lengths in the gray and white matter regions. While the posteriors of the parameters are still spatially varying, the brain tissue segmentation and subject-specific prior are disregarded in this case.
The second approach assumes \textit{piecewise constant posterior} (PCP) distributions of the $D$ and $G$ over the gray and white matter regions. Since the model parameters are no longer spatial fields, the Bayesian calibration in PCP is performed using Delayed-Rejection Adaptive Metropolis sampling algorithm in \cite{dakota} to compute the posterior distributions of $D_{wm}$, $D_{gm}$, $G_{wm}$ and $G_{gm}$. 
Comparing the relative error in NTA and the Dice between the model prediction and imaging data at the testing time points (see Fig. \ref{fig:kde} and Table \ref{tab:prediction}) indicates that the proposed Bayesian framework results in more accurate and more confident computational predictions of tumors than the PCP and SHP (6.66\% and 12.22\% decrease in the means and 200\% and 100\% increase in standard deviations of Dice in \textit{Case I} compared to PCP and SHP, respectively, in \textit{Rat 3}).
%

\section{Conclusions}
\label{sec:conclusions}

We have demonstrated a Bayesian framework for early computational prediction of glioblastoma morphology in individual subjects. The method leverages various MRI modalities to infer the spatial distribution of tumor-specific parameters while quantifying uncertainty in model parameters, measurements, and image segmentation. 
We applied this framework to a reaction-diffusion tumor model and pre-clinical longitudinal MRI data of Wistar rats.
Constructing the subject-specific priors \textit{via} an atlas-based brain tissue segmentation enabled the computational tumor model, constrained by the animal anatomy and the early tumor cellularity imaging data, to provide a robust prediction of the tumor shape (Dice coefficient $>$ 0.88) at later time points. 
The results indicate that accurate and early prediction of tumor shape \textit{via} biophysical models require (i) characterizing tissue heterogeneity by estimating spatially varying model parameters from imaging data and (ii) leveraging prior information about the values and spatial distributions of the parameters.

To the best of our knowledge, this work is the first attempt to quantify uncertainty in the model prediction of the tumor shape and estimate boundary margins. 
Compared to the voxel-wise parameter estimation of tumor growth models (e.g., \cite{ellingson2011}), the proposed framework does not limit the model prediction to the MRI resolution. It also provides more accurate predictions of tumor shape, owing to the spatial correlation of the parameters built into the priors.
Furthermore, the proposed framework can be extended for the early prediction of glioma growth in patients using treatment models. To this end, the cross-validation may be applied to a larger cohort of historical patient data to determine the hyper-parameters.

The basic reaction-diffusion model employed in this work aggregates multiple tumor development mechanisms into the diffusion and proliferation rate parameters. 
Using more complex tumor models that describe additional tumor growth mechanisms (e.g., including 
mechanical deformation \cite{faghihi2020}, 
nutrient distribution \cite{hormuth2020forecasting}, and 
necrosis \cite{roque2017dce}), together with the proposed Bayesian framework, is expected to better uncover the effect of brain tissue heterogeneity on tumor development.
Consequently, additional imaging data, such as 
tissue stiffness map from MRI-Elastography \cite{weis2015}, and 
blood volume fraction estimation from dynamic contrast-enhanced (DCE-)MRI \cite{hormuth2020forecasting} can be readily incorporated into this framework to calibrate tumor-specific parameters of the more advanced models.

In the future, we will extend the proposed method to recursive Bayesian filtering algorithms combined with dynamic control under uncertainty (e.g., \cite{prudencio2015 }) to test enhancement in the subject-specific treatment outcomes. In such an approach, the model is re-calibrated every time additional imaging data is obtained from a subject, and the treatment plans are adaptively updated using the model predicted tumor morphology.

In summary, the proposed image-based inference method and presented results lay the basis for non-invasive and reliable prediction of tumor morphology development and hold great promise to contribute toward clinical prognostic outcomes and personalized treatment.

\section*{Acknowledgment}
This work was supported through funding from the National Cancer Institute
R01CA235800, U01CA174706, CPRIT RR160005 and CPRIT RP220225.
T.E.Y  is a CPRIT Scholar in Cancer Research.

\bibliographystyle{ieeetran} 
\bibliography{refs}

\end{document}